\documentstyle[preprint,aps,prc,epsf]{revtex}
\begin{document}
\preprint{SSF95-12-01}
\draft
\tighten
\title{Photoinduced two-proton knockout and ground-state
 correlations in nuclei}
\author{Jan Ryckebusch \thanks{e-mail : jan@inwfaxp2.rug.ac.be} }
\address{Laboratory for Nuclear Physics, Proeftuinstraat 86, 
B-9000 Gent, Belgium}
\date{\today}
\maketitle
\begin{abstract}
A factorized and analytical 
form for the A($\gamma$,pp) and A(e,e$'$pp) cross section
is proposed.  In the suggested scheme the two-proton knockout cross
sections can be directly analyzed in terms of the ground-state correlation
functions.  Central, spin-spin and tensor correlations are considered.
In the longitudinal channel, the (e,e$'$pp) cross section is predicted
to exhibit a peculiar sensitivity to ground-state correlation effects. 
\end{abstract}
\vspace{0.6 cm}

\vspace{0.6 cm}


\newpage
The nuclear shell-model is a well-established theory for understanding
the structure of atomic nuclei.  Despite its obvious success, for a
long time it has been realized that the strong repulsive nature of the
nuclear force at short internucleon distances is likely to be at the
origin of nuclear effects that are incompatible with the
independent-particle nature of the shell-model.  Obvious signs for
short-range effects have recently been obtained from quasi-elastic
(e,e$'$p) reactions.  Extensive programs at several high-duty electron
facilities predict values for the spectroscopic strength in the
low-energy part of the residual-nucleus spectrum that are consistently
lower than what was expected in the independent-particle model (IPM).
This deviation has been the subject of several investigations, most of
them pointing to short- and long-range correlations to be at the
origin of the ``missing'' spectroscopic strength \cite{mahaux,mutter}.
In analyzing and interpreting the quasi-elastic (e,e$'$p) data it
turned out to be extremely useful to make use of the following
factorized form of the differential cross section \cite{for83} :
\begin{equation}
\frac {d^4 \sigma} {d \epsilon ' d \Omega _{e'} d \Omega _{p} dT _p}
=  E p  \sigma _{ep} \rho( \left| \vec{p}_m \right|, E_x) \; ,
\label{eq:factor}
\end{equation}
where E (p) is the energy (momentum)
of the detected proton and
$\sigma _{ep}$ the elementary cross section for electron scattering on
an off-shell proton.  The spectral function $\rho( \left| \vec{p}_m \right|,
E_x)$ is related to the probability of removing a nucleon with momentum
$p_m$ from the target nucleus and finding the residual nucleus at an excitation
energy $E_x$.  Strictly speaking, the above factorized form for the
(e,e$'$p) cross section is only valid in the plane wave impulse
approximation (PWIA) \cite{for83}, which puts aside effects like final state
interactions (FSI) and photon absorption on two-body currents.

A better understanding of the short-range correlations in nuclei is
believed to come from a profound study of ($\gamma$,pp) and (e,e$'$pp)
reactions.  The underlying idea is that reactions with an
electromagnetic probe which induce two particles to escape are
extremely sensitive to the two-body dynamics inside the target
nucleus.  When the main purpose of the two-nucleon knockout studies is
the short-range part of the two-nucleon dynamics, proton-proton
knockout is to be preferred above proton-neutron knockout.  Indeed,
the latter are also sensitive to correlations of the charge-exchange
type (the most important one being one-pion exchange) that can heavily
mask the effects of short-range nature \cite{cross}.  
With two protons in the final
state one has an enormous freedom when it comes to determining the
position of the hadron detectors.  
Therefore, one could benefit from a simplified form for the cross
section along the lines of Eq.~(\ref{eq:factor}).  For this simplified
form to be useful, one expects it to have some predictive power so
that it can used to map the main sensitivities of the cross sections
and optimize the kinematical conditions.  In this paper we aim at
deriving such a factorized form for the (e,e$'$pp) and ($\gamma$,pp) cross
section.  Factorized expressions for the ($\gamma$,pn) reaction have
been derived in Refs.~\cite{boato,gia91}.  The effect of factorization
was studied in Ref.~\cite{ourqd} and found to be a reasonable
approximation at higher photon energies ($\omega >$150~MeV).  
Therefore, at moderate and larger values of the momentum
transfer, it is to be expected that a factorized form for the
(e,e$'$pp) cross section
provides reasonable estimates. 

\begin{figure}
\centering
\epsfysize=6.cm
\epsffile{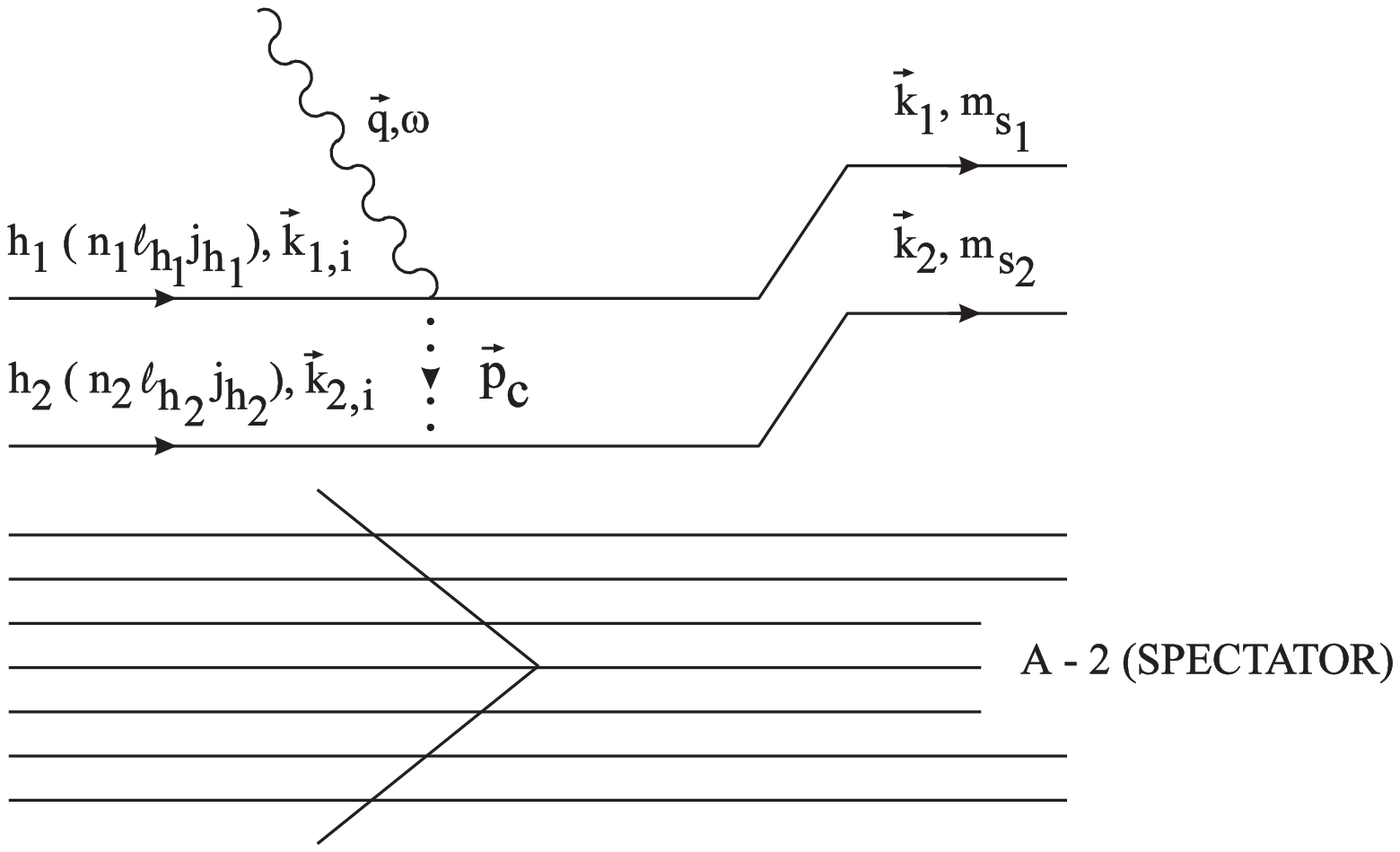}
\caption{The emission of two-protons from a nucleus via (virtual)
photon absorption.}
\label{fig:scheme}
\end{figure}

In order to arrive at a factorized form of a coincidence differential
cross section it is a
common procedure to work in a plane-wave model for the outgoing
nucleons. Furthermore, we adopt the spectator approximation which
means that the residual A-2 nucleons are assumed not to participate
actively in the reaction process.  Similar assumptions are at the
basis of the factorized form (\ref{eq:factor}) for the (e,e$'$p) cross
section. In the two-nucleon knockout case, however, the above
assumptions are not sufficient for the cross section to factorize.  This
can best be explained by considering the two-nucleon knockout process
as schematically depicted in Fig.~\ref{fig:scheme}.  There we assume
that the photon couples to a correlated pair of protons. 
The center-of-mass (com) momentum $ \vec{P}$ of the pair can be
straightforwardly derived and reads 
$ \vec{P} = \vec{k} _{1,i} + \vec{k}_{2,i} = \vec{k} _{1} + \vec{k}_{2} -
\vec{q}$.  The relative momentum $ \vec{p}_{rel} = \left( \vec{k}_{1,i} -
\vec{k}_{2,i} \right)/2 $ on the other hand, will generally depend on
the momentum $\vec{p} _C$ that is exchanged between the two correlated
nucleons.  Thus,
in its most general form the cross section involves an integral over
the $\vec{p} _C$ which prevents it from being written in a
factorized form.  Now, the key approximation for the two-nucleon
knockout cross section to factorize is that  
$\vec{p}_{C} \approx \vec{0}$.
This assumption is
related to the ``quasi-deuteron approximation'' which adopts the view
that dinucleons in finite nuclei are mainly residing in relative S 
states \cite{ourqd,gottfried}.  
Let $\vec{p}_{C}=\vec{0}$,  
then the relative momentum of the pair is determined by 
\begin{equation}
\vec{p}_{rel}  =  \vec{k}_{\pm} = \frac {\vec{k}_{1} -\vec{k}_{2}} {2}  
\pm \frac {\vec{q}} {2}  \nonumber \; ,
\end{equation}
respectively
corresponding with the situation that the photon is absorbed on proton
``1'' ($\vec{k}_{-}$) and ``2'' ($\vec{k}_{+}$).  
With all of the
above assumptions the (e,e$'$pp) cross section can be written in a
factorized form : 
\begin{equation}
\frac {d^5 \sigma} {d \epsilon ' d \Omega _{\epsilon '} d \Omega _{1} 
 d \Omega _{2} dT _{p_{2}}}
       = E_1 p_1 E_2 p_2 \sigma _{epp} \left( 
              k _{+} , k  _{-} ,q        \right)
F_{h_{1},h_{2}}(P) \; ,
\label{eq:faceepp}
\end{equation}
with $ k_{\pm} = \mid \vec{k} _{\pm} \mid$ and $F_{h_{1},h_{2}}(P)$
denoting the probability to find a nucleon pair in the single-particle
states $h_1 (n_1 l_{1})$ and $h_2 (n_2 l_{2})$ with
center-of-mass (com) momentum $\vec{P}$. In the IPM this function
reads :
\begin{equation}
F_{h_{1},h_{2}}(P)= \sum _{m_{1},m_{2}} \left| \int d \vec{R} 
e^{i \vec{P} . \vec{R}}
\phi _{n_{1} l_{1}  m_{1}} (\vec{R}) 
\phi _{n_{2} l_{2}  m_{2}} (\vec{R}) 
\right| ^2 \; ,
\end{equation}
with $\phi _{nlm}$ the single-particle wave function for the orbital
$(nlm)$.  
The elementary cross section
$\sigma _{epp}$ can be considered as the equivalent of the 
$\sigma _{ep}$ in Eq.~(\ref{eq:factor}) and describes
the physics of virtual photoabsorption on a diproton embedded in
the target nucleus. Generally,  $\sigma _{epp}$
will depend on the photoabsorption
mechanisms and the relative motion of the pair.  It is precisely in
the relative motion that the highest sensitivity to the correlation
effects could be expected.  We outline a method to derive an
analytical expression for $\sigma _{epp}$. In doing this, we 
account for dinucleon
correlations that go beyond the IPM.

Adopting the PWA for the outgoing particles one has to evaluate the
following type of matrix elements when calculating two-nucleon
knockout cross sections 
\begin{eqnarray}
M^{h_{1} h_{2} J_{R} M_{R}} _{m_{s_{1}} m_{s_{2}}} & = & 
\int d \vec{r} _1
... \int d \vec{r} _A {\cal A} \left(
e ^{ - i \vec{k}_1 . \vec{r}_1} e ^{ - i \vec{k}_2 . \vec{r}_2} 
\Psi _{A-2} ^{h_1 h_2 J_R M_R}(3,...,A) \right)
\nonumber \\
& & \times   
\left<
1/2 m_{s_{1}}, 1/2 m_{s_{2}} \left| 
\left(
\sum _{i=1} ^{A} J_{\mu}^{[1]}(i) 
+ \sum _{i<j=1} ^{A} J_{\mu}^{[2]}(i,j)                       
\right) \right|
\Psi _i (1,2,...,A) \right> \; ,
\label{eq:mel} 
\end{eqnarray}
where ${\cal A}$ is the anti-symmetrization operator and $m_{s_{1}}$
($m_{s_{2}}$)
the spin of escaping particle 1 (2).  The wave function for the residual
nucleus $\Psi _{A-2} ^{h_1 h_2 J_R M_R}$ is the two-hole state that is
created after knocking two protons out of the target nucleus.  The
operators  $ J_{\mu}^{[1]}$ and $J_{\mu}^{[2]}$ are the one- and
two-body parts of the nuclear current.  As proton-proton emission is
not sensitive to charge-exchange, the major component of the 
two-body operator is the isobaric current.  The $J_{\mu=0}^{[1]}$
is related to the charge density operator $\sum _i G_{E} \left( q ^{\mu}
q_{\mu} \right) e_i \delta
(\vec{r} - \vec{r} _i)$.  The transverse components of the one-body
current  $J_{\mu}^{[1]} (\mu = \pm 1)$ is determined by the 
convection and magnetization current.  The
target wave function in the above expression is written as :
\begin{equation}
\Psi _i(1,2,...,A) = F(1,2,...,A) \psi _i (1,2,...,A) \; ,
\end{equation}
where $\psi _i$ is the IPM wave function and the operator F induces
the correlations.  In general, the operator F has many components,
reflecting the full complexity of the nucleon-nucleon interaction.
Calculations, however, have shown that the major correlation 
effects can
be incorporated by considering an operator of the form 
\cite{pieper,revmod} :
\begin{equation}
F(1,2,... A) = {\cal S} \prod _{i<j=1} ^{A} \left[
                f_C(r_{ij}) + \left( f _{\sigma \tau} (r_{ij})
	\vec{\sigma} _i . \vec{\sigma} _j + f _{t \tau} (r_{ij}) 
S_{ij} \right)
        \vec{\tau} _i . \vec{\tau} _j  \right] \; .
\label{eq:cor}
\end{equation}
The first term accounts for central short-range correlations (commonly
referred to as Jastrow correlations), whereas the
other two induce spin-spin ($\vec{\sigma} _i . \vec{\sigma} _j$) and
tensor ($S_{ij}$) correlations.  The operator $\cal{S}$ is the
symmetrization operator.

In determining $\sigma _{epp}$ we sum over the spins of the escaping
nucleons. Further, instead of considering the contribution to each
individual state we evaluate the cross section for emission out of a
particular shell-model combination $(n_1 l_{h_{1}},n_2 l_{h_{2}})$.  This
means that we compute the averaged integrated cross section for a range of
excitation energies in the A-2 system.  The range of excitation
energies will be rather narrow when hole states close to the Fermi level
are probed and grow wider as one or both nucleons are escaping from a
deeper lying shell.  

One can expand the matrix element (\ref{eq:mel}) in terms of $g
\equiv 1- f_C$, $f_{\sigma \tau}$ and $f_{t \tau}$.  
In doing this we have adopted the
so-called ``single-pair approximation'' (SPA) \cite{orlandini}.  
This procedure is equivalent
with an expansion into first order in $g$, $f_{\sigma \tau}$ and $f_{t
\tau}$, retaining only those terms that contain the coordinates of both
active nucleons.  The SPA was earlier applied in
Ref.~\cite{boato} in the context of ($\gamma$,pn) reactions.
Physically, the SPA is equivalent with multiplying the IPM relative wave
function of the active pair with the correlation operator
of Eq.~\ref{eq:cor}. With all these assumptions one obains
the following general expression for $\sigma _{epp}$ :
\begin{equation}
\sigma _{epp} = \sigma _ M  f_{rec} ^{-1}
\left[ \frac {q _{\mu} ^4} {q ^4} w_L + 
 \left( 
          \frac {-q _{\mu} ^2} {2q ^2} + tan ^2 \frac {\theta _e} {2}
\right)
w_T + \frac {q _{\mu} ^2} {2q ^2} w_{TT} + 
\frac {1} {\sqrt{2}} \frac {q _{\mu} ^2} {q ^3} (\epsilon + \epsilon
') tan \frac {\theta _e} {2} w_{LT} \right] \; ,
\label{eq:seep}
\end{equation}
where $f_{rec}$ is the recoil factor and $\sigma _M$ the Mott cross
section. The w's depend on the current operator and the different terms
in the correlation operator (\ref{eq:cor}).  After lengthy
calculations one arrives at the following analytical expressions : 
\begin{eqnarray}
w_L & = & 4e^2  \left( g(k_+) + g(k_-) \right)^2
\left( G_E(q_{\mu}q^{\mu}) \right) ^2
+ 40 e^2  \left( 
f_{\sigma \tau}(k_+) + f_{\sigma \tau}(k_-) \right)^2
\left( G_E(q_{\mu}q^{\mu}) \right) ^2 
\nonumber \\
& &
+ 24 e^2  \left( g(k_+) + g(k_-) \right) 
\left( f_{\sigma \tau} (k_+) + f_{\sigma \tau} (k_-) \right)
\left( G_E(q_{\mu}q^{\mu}) \right) ^2 
\nonumber \\
& & 
+ \frac {16} {3} \sqrt{ \frac {\pi} {5}} e^2 
\left( g(k_+) + g(k_-) \right) 
\left( f_{t \tau}^0 (- \vec{k} _+) + f_{t \tau}^0 (-\vec{k}_-) \right)
\left( G_E(q_{\mu}q^{\mu}) \right) ^2 
\nonumber \\
w_T & = & \frac {\mu _p ^2 e^2 q^2} {M_p^2} 
\left( g(k_+) - g(k_-) \right)^2 
\left( G_E(q_{\mu}q^{\mu}) \right) ^2  \nonumber \\
   & + & \frac {e^2} {2M_p^2} 
     \left[
          \left( k_{1,x} g(k_-) + k_{2,x} g(k_+) \right)^2
        + \left( k_{1,y} g(k_-) + k_{2,y} g(k_+) \right)^2 
     \right]
\left( G_E(q_{\mu}q^{\mu}) \right) ^2 \nonumber \\
& + &  \frac {256} {81} \left( \frac 
         {f_{\gamma N \Delta} f_{\pi N \Delta} f_{\pi NN}}    
         { m_{\pi} ^3}  \right)^2 
         G_{\Delta} ^2 \left( G_E(q_{\mu}q^{\mu}) \right) ^2 
         \left( \vec{q} \times \left( \frac
                                     {\vec{k}_1 - \vec{k} _2}
	                             {2}
                          \right) \right)^2  \nonumber \\
& & \times \left[ k_+^2 \left( \frac {1} {k_+^2+m_{\pi}^2} \right)^2
+ k_-^2 \left( \frac {1} {k_-^2+m_{\pi}^2} \right)^2
-2 \vec{k}_+ \cdot \vec{k}_- \frac {1} {k_+^2+m_{\pi}^2}
\frac {1} {k_-^2+m_{\pi}^2} \right] \nonumber \\
w_{LT} & = & 0 \nonumber \\
w_{TT} & = & \frac {- 2 \mu _p ^2 e^2 q^2} {M_p^2} 
\left( g(k_+) - g(k_-) \right)^2 \left( G_E(q_{\mu}q^{\mu}) \right) ^2
\nonumber \\
   & - & \frac {e^2} {M_p^2} 
     \left[
          \left( k_{1,x} g(k_-) + k_{2,x} g(k_+) \right)^2
        - \left( k_{1,y} g(k_-) + k_{2,y} g(k_+) \right)^2 
     \right]
\left( G_E(q_{\mu}q^{\mu}) \right) ^2 
\nonumber \\
& - &  \frac {256} {81} \left( \frac 
         {f_{\gamma N \Delta} f_{\pi N \Delta} f_{\pi NN}}    
         { m_{\pi} ^3}  \right)^2 
         G_{\Delta} ^2 \left( G_E(q_{\mu}q^{\mu}) \right) ^2 
         \left[ \left( \vec{q} \times \left( \frac
                                     {\vec{k}_1 - \vec{k} _2}
	                             {2}
                          \right) \right)_x ^2  
- \left( \vec{q} \times \left( \frac
                                     {\vec{k}_1 - \vec{k} _2}
	                             {2}
                          \right) \right)_y ^2 \right]
                             \nonumber \\
& & \times \left[ k_+^2 \left( \frac {1} {k_+^2+m_{\pi}^2} \right)^2
+ k_-^2 \left( \frac {1} {k_-^2+m_{\pi}^2} \right)^2
-2 \vec{k}_+ \cdot \vec{k}_- \frac {1} {k_+^2+m_{\pi}^2}
\frac {1} {k_-^2+m_{\pi}^2} \right] 
\; .
\end{eqnarray}
It should be noted that the tensor and spin-spin correlations have
only been considered for the longitudinal channel $w_L$. 
We have defined the $xz$ plane as
the electron scattering plane with $\vec{q}$ along the z axis.  In
deriving the
isobaric current contribution to the above expression we have used the
operator 
specified in Ref.~\cite{marc} from which also all the coupling
constants are taken.  The
$G_{\Delta}$ is the $\Delta$ propagator in which an energy-dependent
$\Delta _{33}$ has been introduced \cite{marc}.  
The functions $g(k)$, $f_{\sigma \tau}(k)$  and $f_{t \tau}(k)$ 
occuring in the above
expression are the Fourier transforms of the central functions
occurring in the correlation operator (\ref{eq:cor}) :
\begin{eqnarray}
g(p) & \equiv & 
\int d \vec{r} e ^{i \vec{p}. \vec{r}} (1
-f_C(r)) \nonumber \\  
f_{\sigma \tau } (p) & \equiv & 
\int d \vec{r} e ^{i \vec{p} . \vec{r}} f_{\sigma \tau} (r) \nonumber \\
f_{t \tau } ^0 (\vec{p}) & \equiv & 
\int d \vec{r} e ^{i \vec{p} . \vec{r}} Y_{20} (\Omega) 
f_{t \tau} (r) \; .
\end{eqnarray}
In the absence of ground-state correlations only the transverse
$\Delta$ current would give a non-vanishing contribution to
$\sigma_{epp}$.  The longitudinal channel $w_L$ is totally determined
by the ground-state correlations.  

Note that the $\sigma _{epp}$ as written in Eq.~(\ref{eq:seep}) has
formally the same form as the $\sigma _{ep}$ in the (e,e$'$p) case.
The physics situation is, however, very different. In the (e,e$'$p)
case, the $\sigma _{ep}$ contains information on
electron scattering on a bound, off-shell nucleon. In quasi-elastic
kinematics, this scattering process is predominantly sensitive to the
one-body aspects of the target nucleus which are relatively well
understood.  For the (e,e$'$pp) case the situation is completely
different.  Here, $\sigma _{epp}$ contains information on the
two-body
aspects of the nuclear system.  The $F_{h_{1},h_{2}}(P)$ occuring in
the factorized (e,e$'$pp) cross section (\ref{eq:faceepp}) is a 
less-challenging
quantity.  As it deals with the c.o.m. motion of dinucleons it is
rather insensitive to nucleon-nucleon correlations at small and
moderate values of P.  Recent calculations \cite{orlandini} predict
some sensitivity at large momenta P.  However, by then the
function $F_{h_{1},h_{2}}(P)$ becomes so small in absolute magnitude that  
extremely small two-nucleon knockout cross sections can be expected.
More favourable kinematical situations for probing 
correlations are created when considering relatively small values of
P : in those cases larger cross sections are faced and the 
IPM predictions for $F_{h_{1},h_{2}}(P)$ can be considered as
realistic choices.      

\begin{figure}
\centering
\epsfxsize=17.5cm
\epsffile{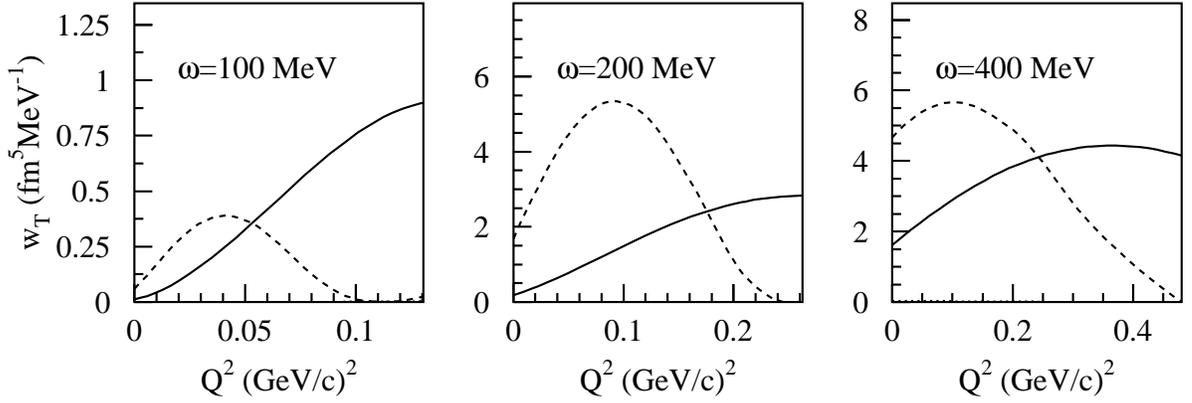}
\caption{The $Q^2$ dependence of the transverse term $w_T$ in $\sigma
_{epp}$ for three values of the energy transfer.  The kinematical
condition is fixed through : $\vec{P}$=50~(MeV)~$\vec{1}_x$, $\theta
_1$=135~$^o$ and $\phi_1$=0~$^o$.  The dotted (solid) line shows the
contribution from photoabsorption on the one-body convection
(magnetization) current.  The contribution from intermediate 
$\Delta _{33}$ production is
shown with the dashed line.  In these calculations a Gaussian central
correlation function was used.  See text for further details}
\label{fig:qdepen}
\end{figure}

The $(\gamma,pp)$ cross section is uniquely
determined by the $w_T$ term and reads :
\begin{equation}
\frac {d^3 \sigma} {d \Omega _{1} 
 d \Omega _{2} dT _{p_{2}}}
       = E_1 p_1 E_2 p_2 \sigma _{\gamma pp} \left( 
              k _{+} , k  _{-} ,q_{\gamma}        \right)
F_{h_{1},h_{2}}(P) \; ,
\end{equation}
with,
\begin{equation}
\sigma _{\gamma pp} \left( 
              k _{+} , k  _{-} ,q_{\gamma}
                    \right) =
\frac {1} {2 E_{\gamma} (2 \pi) ^5} f_{rec} ^{-1}  w_T \; .
\end{equation}
In the transverse response function $w_T$ 
the contribution from the isobaric current and the terms
related to ground-state correlations are competing.  
The $Q^2=-q_{\mu} q^{\mu}$ dependence
of the different terms in $w_T$ is investigated in
Fig.~\ref{fig:qdepen}.  We have considered in-plane kinematics (which
means that both protons are escaping in the electron scattering plane) 
with a fixed $\vec{P}$ (which is chosen to point along the x axis), 
$\theta _{1}$ and $\phi _{1}$ (polar and azimuthal angle of escaping
proton 1).  For the results of Fig.~\ref{fig:qdepen} the central
correlation function of the Gaussian type $g(r)= 0.51 e^ {1.52
(fm^{-2}) r^2}$  
as suggested in
Ref.~\cite{co} was used.  For all
three energy transfers considered, the $\Delta$(1232) current produces
the largest contribution at the real photon point ($Q^2$=0).  Accordingly,
($\gamma$,pp) reactions are predicted to exhibit
a rather small sensitivity to Jastrow correlations.  The
second term in $w_T$, which corresponds with photoabsorption on the
one-body convection current, has a marginal effect on the cross
section.  The term related to the 
magnetization current (first term in the 
expression for $w_T$) has a $q^2$ dependence which makes it a
dominant contribution at higher $Q^2$.    

\begin{figure}
\centering
\epsfxsize=17.5cm
\epsffile{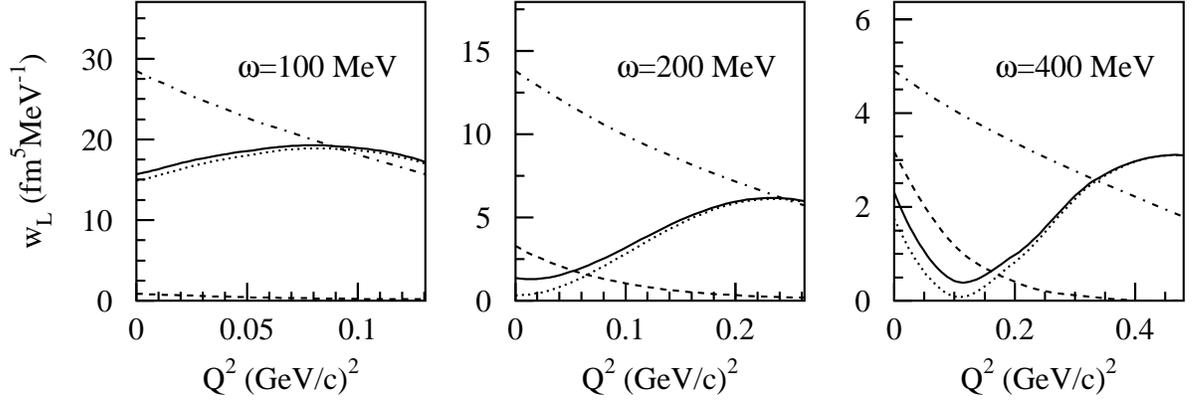}
\caption{As in Fig.~\protect{\ref{fig:qdepen}} but now for the
longitudinal term $w_L$.  The dashed curves only include Jastrow
correlations, the dotted curves include Jastrow and spin-spin
correlations, and finally the solid curves include Jastrow, spin-spin
and tensor correlations.  All these results have been obtained with
the variational correlation functions of Ref.~\protect{\cite{revmod}}.  The
dot-dashed line includes only Jastrow correlations, but now calculated
with the Gaussian correlation function of Ref.~\protect{\cite{co}}.}
\label{fig:qdepen_2}
\end{figure}

In Fig.~\ref{fig:qdepen_2} the $Q^2$ dependence of the longitudinal
term ($w_L$) in
$\sigma _{epp}$ is
displayed for the same kinematical conditions as for
Fig.~\ref{fig:qdepen}. 
The first and second term in $w_L$ are
originating from the central and spin-spin correlations respectively.
The interference between the spin-spin and central correlations gives
rise to the third term of $w_L$. We remark that within the adopted
approximations the contribution from the tensor correlations to 
$w_L$ is restricted to an interference term with the
central correlations.  
For the $w_L$ we have
investigated the role of the different terms in the correlation
operator (\ref{eq:cor}).  Whereas central Jastrow correlations are
frequently addressed in the literature, studies that include  
also spin-spin and tensor correlations are far less abundant.
Using variational techniques the functions 
$f_C, f_{\sigma \tau}$  and $f_{t \tau}$ have been determined 
starting from a 
realistic nucleon-nucleon interaction
\cite{pieper,revmod}.
We have used the correlation functions from 
the $^{16}$O calculations of Ref.~\cite{pieper}.  
It is striking that the Jastrow
correlation function $f_C$ that comes out of these variational  
methods  is very soft.  For the sake of reference, we show in
Fig.~\ref{fig:qdepen_2} also the result when including only the
Gaussian Jastrow correlation function that was introduced earlier.
With this choice of the central correlation function the longitudinal
(e,e$'$pp) strength is predicted to be considerably larger than the strength
produced by the variational central correlation function $f_C$.  The
variational method, however, is a more complete model in the sense
that it accounts for the full complexity of the ground-state correlations.  
It is clear from 
Fig.~\ref{fig:qdepen_2} that spin-spin correlations are predicted to
contribute 
substantially to the longitudinal (e,e$'$pp) strength.  The effect of
tensor correlations is rather marginal.  This is not too surprising as
they are generally considered to be a predominant proton-neutron
correlation effect \cite{leidemann}.  Note that the predictions for
$w_L$  depend dramatically on the model assumptions with respect to
the ground-state correlations.  

\begin{figure}
\centering
\epsfysize=15.cm
\epsffile{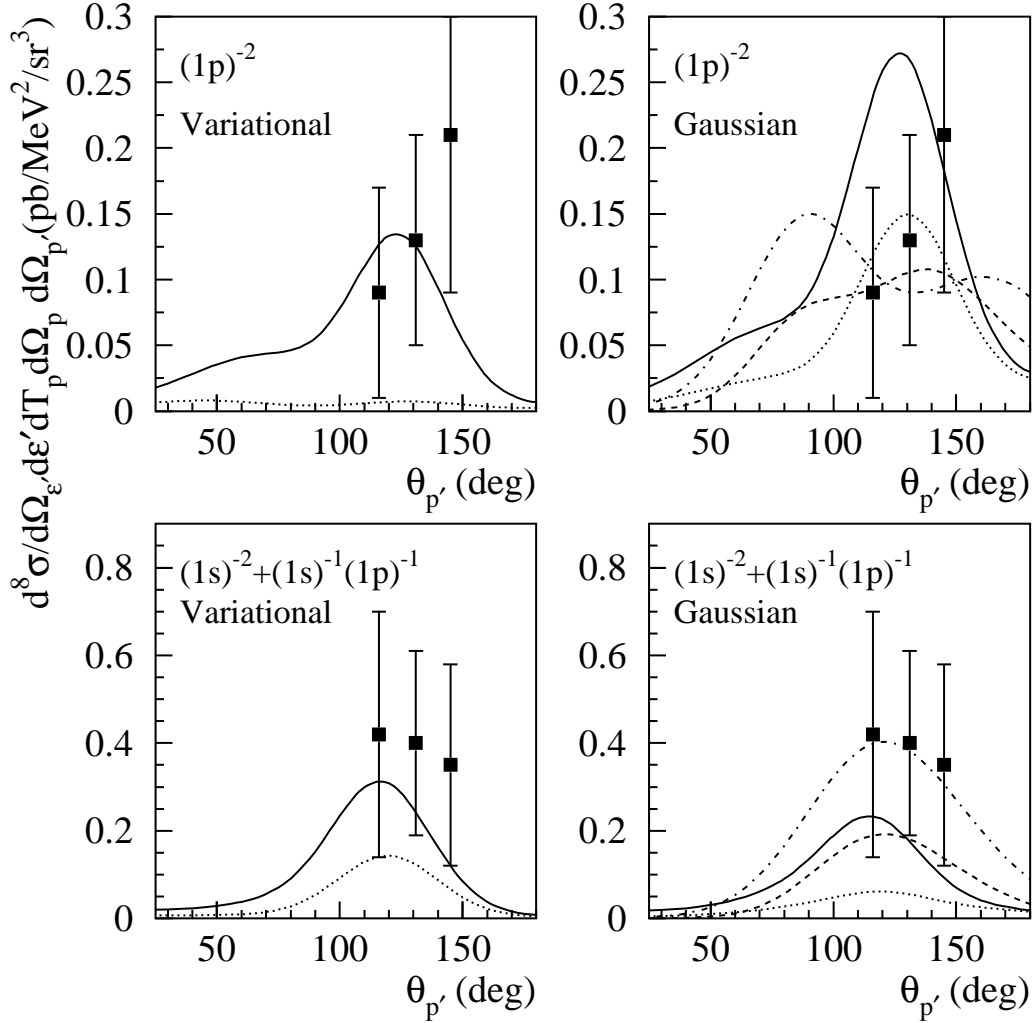}
\caption{The $^{12}$C(e,e$'$pp) cross section for $\epsilon$ =
475~MeV, $\omega$=212~MeV and q=270~MeV/c. One of the proton
scattering angles was fixed at 27$^{o}$.  The dotted line is
the calculated contribution from the ground-state correlations
and the solid line is the prediction when including both ground-state
correlations and intermediate $\Delta _{33}$ creation.
Results for two types of correlation functions are shown.
  The data are from
Ref.~\protect{\cite{leon}}. The dot-dashed (dashed) line shows
the result of an unfactorized calculation with plane (distorted)
outgoing proton waves.}
\label{fig:data}
\end{figure}

In Fig.~\ref{fig:data} we compare the predictions of the suggested
factorized cross section with recent $^{12}$C(e,e$'$pp) 
data \cite{leon} and with unfactorized calculations.  
Two different regions in the excitation spectrum
of the A-2 system were considered : $E_x \leq $~13~MeV and
23~MeV$ \leq E_x \leq$~48~MeV.  The first region can 
be attributed to $(1p)^2$
knockout and the second to both  $(1p)(1s)$ and $(1s)^2$ knockout.
For both of these regions we have calculated the angular cross
sections with the variational and Gaussian correlation functions.   
With both choices one gets cross sections that are compatible with the
data. 



For the Gaussian results we have compared the predictions of the
factorized approach to the results of an unfactorized
calculation.  These calculations are based on an extension of the model
outlined in Refs.~\cite{jan1,jan2} and
involve the same physics components as the
unfactorized approach in the sense that both central correlations and
$\Delta _{33}$ effects are considered.  Unlike in a factorized model,
one is no longer bound to a plane wave description for the outgoing
particles in an unfactorized approach \cite{jan1,pavia1,babib}.
The right panels of
Fig.~\ref{fig:data} show the results of the unfactorized calculations
with both plane and distorted outgoing proton waves. Consequently the
sole difference between the plane wave unfactorized calculations and
the analytical approach discussed above is the treatment of the
relative motion of the initial pair. It is noted that the unfactorized
calculations produce angular cross sections that are wider than the
analytical predictions.  This is not too surprising given the fact
that the relative motion of the initial diproton is now handled in its
full complexity.  After all, the factorized predictions could be
considered as reasonable given that they can be performed in just a
fraction of the computing time that the unfactorized calculations
consume.  Eventhough the analytical expression
$\sigma _{epp}$ should not be considered as a fully-fledged
alternative for the cumbersome unfactorized calculations, it could
help in optimizing the kinematical conditions and determining
the major trends and sensitivities of the cross section.

Summarizing, we have derived a factorized form for the (e,e$'$pp) and 
($\gamma$,pp) cross section.  This implies rather
simple analytical expressions that should give a
more transparent handle on the different physics components of
photoinduced two-proton knockout reactions.  Within the 
method developed here we have 
illustrated the sensitivity of the (e,e$'$pp) cross
sections to ground-state correlation effects.  

{\em Acknowledgement} This work has been supported by the National
Fund of Scientific Research (NFWO). The author is grateful to Steven
Pieper for kindly providing him with the correlation functions from
the variational calculations.

\end{document}